# Large in-plane deformation of RuO$_6$ octahedron and ferromagnetism of bulk SrRuO$_3$

Sanghyun Lee[1,2,3], J. R. Zhang[4], S. Torii[4], Seongil Choi[1,2,3], Deok-Yong Cho[1], T. Kamiyama[4], Jaejun Yu[2,5], K. A. McEwen[6], and Je-Geun Park[1,2,5,7#]

[1] Center for Correlated Electron Systems, Institute for Basic Science, Seoul National University, Seoul 151-747, Korea
[2] Center for Strongly Correlated Materials Research, Seoul National University, Seoul 151-742, Korea
[3] Department of Physics, SungKyunKwan University, Suwon 440-746, Korea
[4] Institute of Materials Structure Science & J-PARC Center, KEK, Tsukuba 305-0801, Japan
[5] FPRD, Department of Physics & Astronomy, Seoul National University, Seoul 151-747, Korea
[6] Department of Physics & Astronomy, and London Centre for Nanotechnology, University College London, London WC1E 6BT, UK
[7] Center for Korean J-PARC Users, Seoul National University, Seoul 151-742, Korea

SrRuO$_3$ is a ferromagnetic metal with several unusual physical properties such as zero thermal expansion below T$_c$, so-called Invar behavior. Another anomalous feature is that the a-axis lattice constant is larger than the b-axis lattice constant, a clear deviation from the predictions of the Glazer structural description with rigid RuO$_6$ octahedron motion. Using high resolution neutron diffraction techniques, we show how these two structural anomalies arise from the irregular in-plane deformation, i.e. plastic behavior of the RuO$_6$ octahedron, a weak band Jahn-Teller distortion. We further demonstrate that the ferromagnetic instability of SrRuO$_3$ is related to the temperature-induced localization of Ru *4d* bands.

# I. Introduction

Over the past three decades, transition metal oxides have been at the center of strongly correlated electron physics with numerous examples including high temperature cuprates, colossal magnetoresistance (CMR) manganites, multiferroics, etc. By virtue of their strong correlations, every aspect of these oxides, be it electronic, magnetic or structural, is interlinked. Frequently, one of these order parameters holds the key to a proper understanding of the otherwise unrelated features. A prime example is the metal-insulator and structural transitions accompanying the orbital ordering of the CMR manganites [1].

Oxygen octahedrons ($O_6$) are the elementary building blocks of these oxide materials including perovskite, pyrochlore, and spinel structures. The so-called Glazer angles, defined as the rotating & tilting angles of the oxygen octahedron along the three principal crystallographic axes, can describe the most common forms of structural distortions, for example those observed in the perovskite oxides [2]. The underlying basic assumption behind the Glazer description of the structural transitions is that the oxygen octahedron is a rigid object without its own irregular distortions. One natural consequence of the Glazer description for the orthorhombic Pbnm space group is that the a-axis should always be shorter than the b-axis. This rule is found to hold for most orthorhombic perovskite compounds except for few notable exceptions such as $SrRuO_3$, whose a-axis is not shorter, but longer than its b-axis. This anomaly has remained a puzzle over the past years, and it was theoretically predicted to be related to the irregular deformation of the oxygen octahedron [3].

$SrRuO_3$ belongs to the layered Ruddlesden-Popper series of $Sr_{n+1}Ru_nO_{3n+1}$ compounds and has the simplest structure of all with n=$\infty$, i.e. a perovskite structure [4]. Due to its good metallic properties as well as favorable lattice matches with several important oxide materials, it has been extensively used as a universal electrode in the thin film community [5]. Despite its simple structure, $SrRuO_3$ has an unusually high Curie temperature of $T_c$=165 K, with a large ordered magnetic moment of over 1 $\mu_B$/Ru, and exhibits very unusual zero thermal expansion in the ferromagnetic state, so-called Invar behavior. $SrRuO_3$ is one of the few known ferromagnetic Ru oxides: other examples include $Sr_4Ru_3O_{10}$ [6]. However, as far as we are aware, none of the other ferromagnetic Ru oxides show the unusual volume anomaly of $SrRuO_3$ below their respective ferromagnetic transitions.

Although the first reports of the ferromagnetism of $SrRuO_3$ appeared in the literature many years ago [7,8] and extensive studies have since been made [9-13], its origin and, more importantly, the relation between the ferromagnetic ground state and the zero thermal expansion behavior still remains far from understood. Moreover, it is particularly significant that although $CaRuO_3$ has a very similar crystal structure, it exhibits neither magnetic ordering nor Invar behavior.

In this paper, we report a detailed experimental study of the temperature evolution of the irregular deformation of the oxygen octahedron in $SrRuO_3$, by using the state-of-the-art high resolution powder diffractometer (S-HRPD) recently built at J-PARC. We clearly demonstrate that this lattice anomaly is due to an extremely large (2%) in-plane deformation of $SrRuO_3$, in stark contrast with $CaRuO_3$ that is more consistent with the Glazer picture of rigid octahedron rotations. We also show how the deformation and the volume change of the $RuO_6$ octahedron are related to the ferromagnetic instability of $SrRuO_3$.

## 2. Experimental details

SrRuO$_3$ forms in an orthorhombic structure (Pbnm space group) with two oxygen sites, and its RuO$_6$ octahedron has three kinds of Ru-O bonds: two in-plane Ru-O bonds (short Ru-O1 & long Ru-O1 bonds) and a third bond (Ru-O2) along the c-axis. As it is well known that the ferromagnetic transition temperature of SrRuO$_3$ is rather sensitive to sample quality [12,13], we have optimized our synthesis procedures by monitoring the quality with the aid of subsequent x-ray diffraction studies and magnetization measurements. We prepared high quality powder samples by a solid state reaction method with starting materials of SrCO$_3$ and CaCO$_3$ of 99.995 % purity and RuO$_2$ of 99.9 % purity. We calcinated the pelletized starting materials at 900 °C for 24 hours. After further grinding, they were subjected to another heat treatment at 1300 °C for 24 hours followed by a controlled cooling at a rate of 1 °C/min. In the case of SrRuO$_3$, it was found that a single sintering process was better than repeated heat treatments, whereas our best CaRuO$_3$ sample was obtained after repeating the final heat treatment for three times. We monitored the quality of the samples by measuring x-ray diffraction using both a commercial high resolution powder diffractometer (Bruker AXS D8 FOCUS) and a high resolution powder diffractometer at the Pohang Accelerator Laboratory, Korea.

We also measured the resistivity by using a home-made set-up, and the magnetization under an applied field of 500 Oe from 2 to 300 K by a SQUID magnetometer (MPMS-5XL, Quantum Design, USA). The high quality of our samples was confirmed by the fact that we achieved a very high Curie temperature of $T_c$=162 K in the measurements of resistivity, magnetization, and neutron diffraction, which is very close to that reported for single crystals [14] (see Fig 1).

High resolution ($\Delta d/d \cong 3.5 \times 10^{-4}$) neutron powder diffraction experiments were carried out from 10 to 300 K using the S-HRPD beamline at the J-PARC Center, Japan. For the magnetic structure determination, additional neutron diffraction experiments were carried out with a neutron wavelength of 1.835 Å using the HRPD powder diffractometer at the HANARO reactor, Korea. For each measurement, we used about 4 g of powder samples and employed the Rietveld refinement program, Fullprof, to analyze the data [15]. Our refinement results are summarized for three representative temperatures and the data taken 10 K are shown for both SrRuO$_3$ and CaRuO$_3$ in Fig. 2. The neutron diffraction studies revealed that the ordered magnetic moment of Ru is 1.4 $\mu_B$/f.u. We note that there are wide variations in the reported values of the magnetic moments ranging from 0.9 to 1.5 $\mu_B$ [4,7,16] probably due to an incomplete saturation of the moments or problems of sample quality. Concerning with the easy axis, we found that our refinement results prefer the moment pointing along the b-axis although there are small variations in the goodness of the fit even if we put the moment along the two other axes of the orthorhombic structure.

### 3. Results and Analysis

Using these powder samples of SrRuO$_3$ and CaRuO$_3$, we carried out further higher resolution neutron powder diffraction experiments at the S-HRPD beamline. As shown in Fig. 3a, the three lattice constants (a, b, and c) of SrRuO$_3$ contract upon cooling, until below the transition temperature both the b- and c-lattice constants appear to be *temperature independent* while the a-lattice constant exhibits normal behavior. This unusual thermal expansion of both the b & c axes below the transition can also be seen in the temperature dependence of the unit cell volume in Fig. 3a.

For further analysis, we used a standard theoretical Debye-Grüneisen formula for thermal expansion. In the Debye-Grüneisen formula, the temperature dependence of the unit cell volume is described by $V(T) = V_0 \left[1 + \frac{E(T)}{Q - bE(T)}\right]$, where $V_0$ is the unit cell volume at zero temperature, $Q = (V_0 B_0/\gamma)$, and $b = (B'_0 - 1)/2$. $B_0$ is the zero temperature isothermal bulk modulus with $B'_0$ being its first derivative with respect to pressure and γ the thermal Grüneisen parameter: $B_0$ and $B'_0$ is experimentally determined to be 192 GPa and 5 for SrRuO$_3$, respectively [17]. The internal energy due to lattice vibrations, $E(T)$, is then given by the Debye model: $E(T) = \frac{9nk_BT}{(\theta_D/T)^3} \int_0^{\theta_D/T} \frac{x^3}{e^x-1} dx$, where $\theta_D$ is the Debye temperature, *n* the number of atoms per unit cell, and $k_B$ the Boltzmann constant.

The theoretical curve (lines in Fig. 3a) was calculated by using the Debye-Grüneisen formula with the following set of parameters for SrRuO$_3$: $\theta_D$= 526 K, V$_0$= 241 Å$^3$, Q = 3.18×10$^{-17}$ J, and b = 2. The value of b is fixed at 2 to be consistent with the known $B'_0$ = 5 value of SrRuO$_3$ [17], and this set of parameters gives γ (=$V_0B_0/Q$) of 1.46 for SrRuO$_3$. Slightly different parameters could well produce an equally good fitting to the data as shown for other oxides [18]. We note that the deviation seen below T$_c$ in the thermal expansion, the Invar behavior, has been previously reported [19]. However, the microscopic origin of the structural anomaly, the unmistakable hallmark of the ferromagnetism in SrRuO$_3$, remains unanswered. On the other hand, CaRuO$_3$ does not show any anomaly in the temperature dependence of the structural parameters (see Fig. 3a). The temperature dependence of the unit cell volume of CaRuO$_3$ can be well described by the following parameters: $\theta_D$= 532 K, V$_0$= 226 Å$^3$, Q= 3.04×10$^{-17}$ J, and b = 2. This set of parameters and using the bulk modulus $B_0$ = 192 GPa of SrRuO$_3$ gives us the γ value of 1.43 for CaRuO$_3$, which is similar to the γ value of SrRuO$_3$.

In order to understand the anomalous behavior of SrRuO$_3$ at an atomic & microscopic level, we have carefully examined its crystal structure, revealing previously unreported details of the structural changes occurring below T$_c$. First, we discovered a significant, and previously unrecognized, temperature dependence of the three Ru-O bonds for SrRuO$_3$ (see Fig. 3b). For comparison, we have also plotted the data for CaRuO$_3$ together. In order to avoid any systematic errors in our studies, we took data for three successive thermal cycles: these were all in good agreement. Our results demonstrate that upon cooling the short Ru-O1 bond exhibits a sharp increase, while the long Ru-O1 bond decreases slightly and Ru-O2, the shortest one of the three, remains almost temperature independent with a weak anomaly at T$_c$. This temperature dependence of the Ru-O bonds of SrRuO$_3$ is in striking contrast to that of CaRuO$_3$ with apparently weaker temperature dependence although it exhibits more distorted Ru-O bonds. We note that the larger error bars of the CaRuO$_3$ data are due to the relatively poor statistics of the raw data compared with those of SrRuO$_3$. Unlike the case of SrRuO$_3$, the two in-plane Ru-O1 bonds of CaRuO$_3$ are much larger than, and well-separated from, the apical Ru-O2 bond (see Fig. 3b).

Using these data, we calculated the Glazer rotation ($\Phi_R$) and tilting ($\Phi_T$) angles as defined along the [001] and [110] axes of the cubic perovskite [3]. For our calculations, we modified the [001] rotation angle by taking the O1-O1-O1 angle along the orthorhombic b-axis instead of O1-O1-O1 on the ab plane as used in Ref. 3, because the O1 atoms along the b-axis have the same z position while the O1 atoms separated half way along the a-axis do not (see Fig. 4 for the crystal structure). As one can see in Figs. 3c and 3d, the Glazer angles are larger by about 70% for CaRuO$_3$ than for SrRuO$_3$, consistent with the view that CaRuO$_3$ has more distorted Ru-O-Ru links, and so a supposedly narrower bandwidth compared with SrRuO$_3$. These strongly distorted Ru-O-Ru links due to the larger Glazer angles would then lead to larger resistivity values for CaRuO$_3$ and so driving it closer to a Mott-Hubbard metal-insulator transition as seen by optical studies [11].

However, a closer inspection of several other structural parameters further brings to light that there are more subtle, yet important anomalies. For example, the difference between the two values of both in-plane bond angles and edge lengths of the $RuO_6$ octahedron are bigger for $SrRuO_3$ than for $CaRuO_3$. This unexpected feature is well captured in the plot of O1-O1 distances. Therefore, despite the relatively smaller Glazer angles for $SrRuO_3$ (see Fig. 3c and 3d), the actual in-plane O1-O1 edges are much more deformed for $SrRuO_3$, compared with $CaRuO_3$. This surprising finding has its origin in the O1-Ru-O1 angles. As shown in Fig. 5, (O1-Ru-O1)$_a$ and (O1-Ru-O1)$_b$ angles are 91.08 and 88.92° for $SrRuO_3$ while they are 90.20 and 89.80° for $CaRuO_3$. Therefore, $CaRuO_3$ with a much bigger Glazer rotation and tilting angles involves a more regular in-plane $RuO_6$ octahedron, while an irregular in-plane deformation of about 2% is the unique structural feature of $SrRuO_3$. This surprising result indicates the remarkable plastic behavior of the $RuO_6$ octahedron in $SrRuO_3$ as opposed to a more accepted view of a rigid octahedron for other oxides.

In order to demonstrate how these observed structural anomalies are reflected in the lattice constants, we have calculated their temperature dependence using the following formulae: $a=2a_{Oct}\cos(\Phi_R)\cos(\Phi_T)$, $b=2b_{Oct}\cos(\Phi_R)$, and $c=4d_{Ru-O2}\cos(\Phi_T)$, where $a_{Oct}$ and $b_{Oct}$ represent the octahedral edges along the a- and b-axes, while $d_{Ru-O2}$ is the Ru-O2 bond length. To highlight the effects of the irregular $RuO_6$ deformation, we have specifically calculated the b-lattice constant using three methods as shown in Fig. 5: first, (square) with the mean Ru-O1 bond length fixed at an experimental room temperature value; second, (triangle) with the measured temperature-dependent <Ru-O1> bond length fully considered; last, (diamond) with the measured temperature-dependent O1-O1 edge lengths so taking into account the full deformation. For all three cases, we used the measured temperature-dependent Glazer angles in our calculations. Two things are noteworthy in Fig. 5: first, the unusual temperature anomaly below $T_c$, the Invar anomaly, is only captured correctly in our calculations when we used the measured temperature-dependent <Ru-O1> bond (or O1-O1 edge) length (triangle and diamond symbols). Second, the calculated b lattice constants are always larger than the measured a lattice constant unless we consider the measured in-plane deformation correctly (diamond symbols). This observation once again highlights the important role played by the $RuO_6$ irregular deformation as regards the structural anomalies. To further illustrate the effects of the $RuO_6$ deformation, we have calculated all three lattice constants of $SrRuO_3$ and $CaRuO_3$ by considering the temperature-dependent Glazer angles both with and without the $RuO_6$ deformation (see Fig. 6 for the summary). Notice that even the abnormal temperature dependence of the b-axis lattice constant of $CaRuO_3$ is due to the $RuO_6$ irregular, albeit weak, deformation. This analysis of ours has certain implications of wider reach that other Pbnm orthorhombic oxides having the a-axis longer than the b-axis are expected to have an irregular deformation of their $O_6$ octahedron like the one observed in $SrRuO_3$.

A further salient feature is the temperature dependence of the average <Ru-O> bond length (see Fig. 7a). The mean <Ru-O> bond length of $CaRuO_3$ decreases in accordance with the theoretical curve with parameters similar to those used for the lattice constants. However, <Ru-O> of $SrRuO_3$ not only deviates from the theoretical curve but also increases as indicated by the dashed line. The increase in <Ru-O> of $SrRuO_3$ coincides with the emergence of the ferromagnetic state at low temperatures. This is in marked contrast to $La_{0.75}Ca_{0.25}MnO_3$, an archetypal CMR compound with 3d electrons, which displays a significant *drop* in the <Mn-O> bond length upon entering its ferromagnetic state [20]. Although we acknowledge that $La_{0.75}Ca_{0.25}MnO_3$ may not be a typical 3d ferromagnet, at least it can be used as a good reference material with the similar behavior: i.e. an oxide with a ferromagnetic phase transition with a large volume anomaly at $T_c$. This increased <Ru-O> bond length and so the volume of the $RuO_6$ octahedron found for $SrRuO_3$ is quite different from those seen in the 3d magnetic systems. Therefore this result provides clear experimental evidence that a different mechanism of ferromagnetism is at work for $SrRuO_3$ compared with (La,Ca)$MnO_3$. Our observation of the irregular deformation in SrRuO$_3$ points towards an interesting possibility of the underlying orbital degree of freedom involved, some kind of weak band Jahn-Teller distortions for SrRuO$_3$, although it is not yet fully stabilized as a long range orbital order as suggested by recent LDA+U band calculations [21].

Naturally, with the bigger <Ru-O> bond length at lower temperatures, the Ru *4d* orbital becomes more localized in the ordered phase, favoring a magnetic ground state. To prove this point, we have calculated the bandwidth (W) of the three compounds using the following empirical formula [22]: $W \sim \frac{\cos\omega}{d_{TM-O}^{3.5}}$, where ω is the tilting angle and d$_{TM-O}$ is a TM(transition metal element)-O bond length (see Fig. 7c). As can be seen, there is a persistent temperature-induced narrowing of the Ru 4d bands for SrRuO$_3$, in contrast to both CaRuO$_3$ and La$_{0.75}$Ca$_{0.25}$MnO$_3$. We note also that there is a close correlation among the three physical quantities of both structural and magnetic origin as shown in Fig. 7b. It further reinforces our view of the microscopic link between the structural anomalies reported here and the ferromagnetic instability.

## 5. Discussion and Summary

As we discussed earlier, the two different magnetic ground states have been found in CaRuO$_3$ and SrRuO$_3$ although both materials have the same crystal structure. Our experimental observations offer some qualitative answers on the origin of the different magnetic ground states. First of all, CaRuO$_3$ has a smaller unit cell volume compared with SrRuO$_3$, which leads to a larger distortion of RuO$_6$ octahedron. This will then introduce stronger *pd* or *dd* hybridization, which itself would be detrimental to stabilizing a magnetic ground state for CaRuO$_3$. On the other hand, SrRuO$_3$ has a relatively larger unit cell volume with somewhat localization tendency, which can be made stronger upon cooling due to the unusual in-plane deformation as we observed. This effect is sufficient enough to tip the balance in favour of a magnetic, in this case ferromagnetic ground state for SrRuO$_3$.

The octahedron distortion we found may be strongest among 4d transition metal oxides, but it is not unique to SrRuO$_3$ alone. In fact, somewhat general discussion on similar behaviour has been given for 3d transition metal oxides in Ref. 24. The fact that Sr has a larger ionic radius than Ca with an ionic radius 1.18 and 1 Å for Sr and Ca, seems to be consistent with the arguments given for 3d transition metal oxides in Ref. 23. However, we note that SrRuO$_3$ with the ionic radius lager than 1.11 Å undergoes an orthorhombic to tetragonal phase transition before becoming a cubic perovskite phase [24] without an intervening rhombohedra R-3c phase, unlike what is discussed in Ref. 23.

Furthermore, it will be interesting to examine whether the in-plane deformation we found in SrRuO$_3$ also exists for other oxides of Pbnm space group with a-axis longer than b-axis such as LaCrO$_3$ [25] and LaGaO$_3$ [26]. More generally, it is interesting to note that irregular deformation may not be that uncommon among transition metal oxides if we can examine them using high resolution instruments as we have found in SrRuO$_3$ in this study and also in multiferroic (Y,Lu)MnO$_3$ in a previous study [27]. In fact, this delicate deformation and the plasticity of the metal-oxygen building block may hold a key to some of unusual properties often found in transition metal oxide materials.

In conclusion, by using a high resolution neutron diffractometer, we have succeeded in unraveling the previously unreported structural details of SrRuO$_3$: namely, the large in-plane deformation of the RuO$_6$ octahedron. By taking into account the measured in-plane deformation of a weak band Jahn-Teller type correctly, we have succeeded in explaining the two key structural anomalies: first, the a-axis being longer than the b-axis, and second, the Invar anomaly. We have also illustrated how the microscopic atomic Invar behavior makes otherwise broad Ru *4d* bands more *localized*, favoring the ferromagnetic ground state.

**Acknowledgements**

We acknowledge S. Lee, C. H. Lee, M. Ranot, D. C. Ahn, and N. S. Shin for technical assistance, and S. Streltsov, J. Y. So, D. T. Adroja, A. Pirogov, C-H. Kim, A. Murani, Y. Endoh and J. P. Attfield for useful discussions. This work was supported by the Research Center Program of IBS (Institute for Basic Science, Grant No. EM1203) in Korea and the National Research Foundation of Korea (Grant No. R17-2008-033-01000-0).

Fig. 1 (Color online) Bulk resistivity and susceptibility are shown with the ordered magnetic moments obtained from neutron powder diffraction data. The vertical dashed line indicates the ferromagnetic transition. The line in (c) is a theoretical curve of mean field type. Insert in (a) shows the temperature derivative of the resistivity (dρ/dT), the peak of which is defined as our ferromagnetic transition temperature.

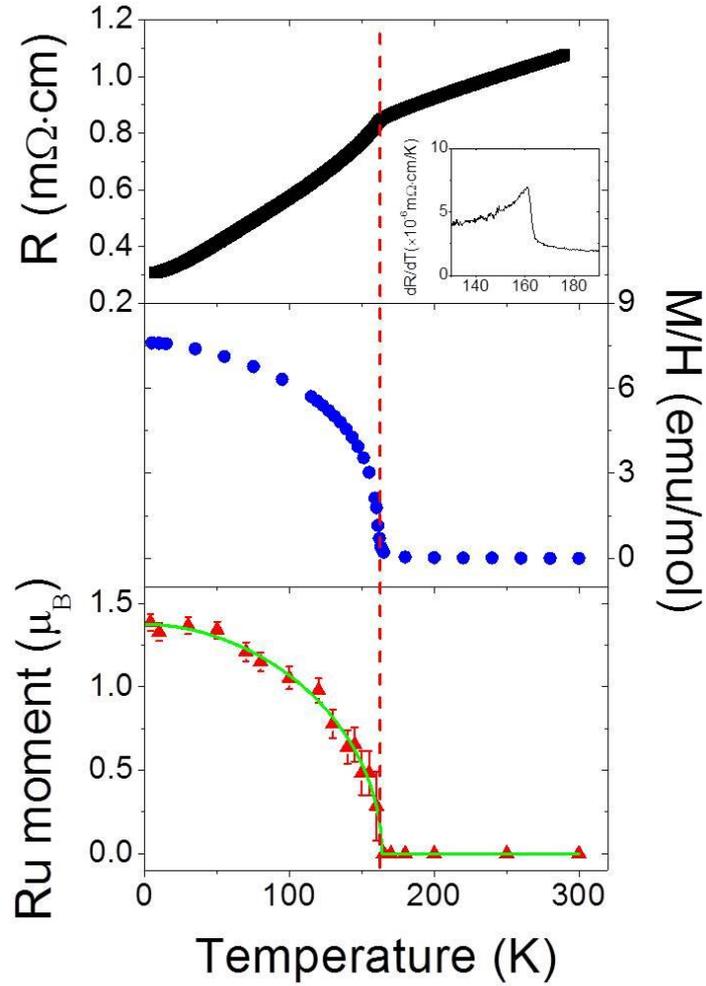

Fig. 2 (Color online) Observed (circle) and calculated (line) neutron diffraction patterns for SrRuO$_3$ and CaRuO$_3$ at 10 K. The lines at the bottom of each figure are the difference curves between the observed and calculated diffraction patterns. The bars indicate the position of the nuclear (top) and magnetic (bottom) Bragg peaks.

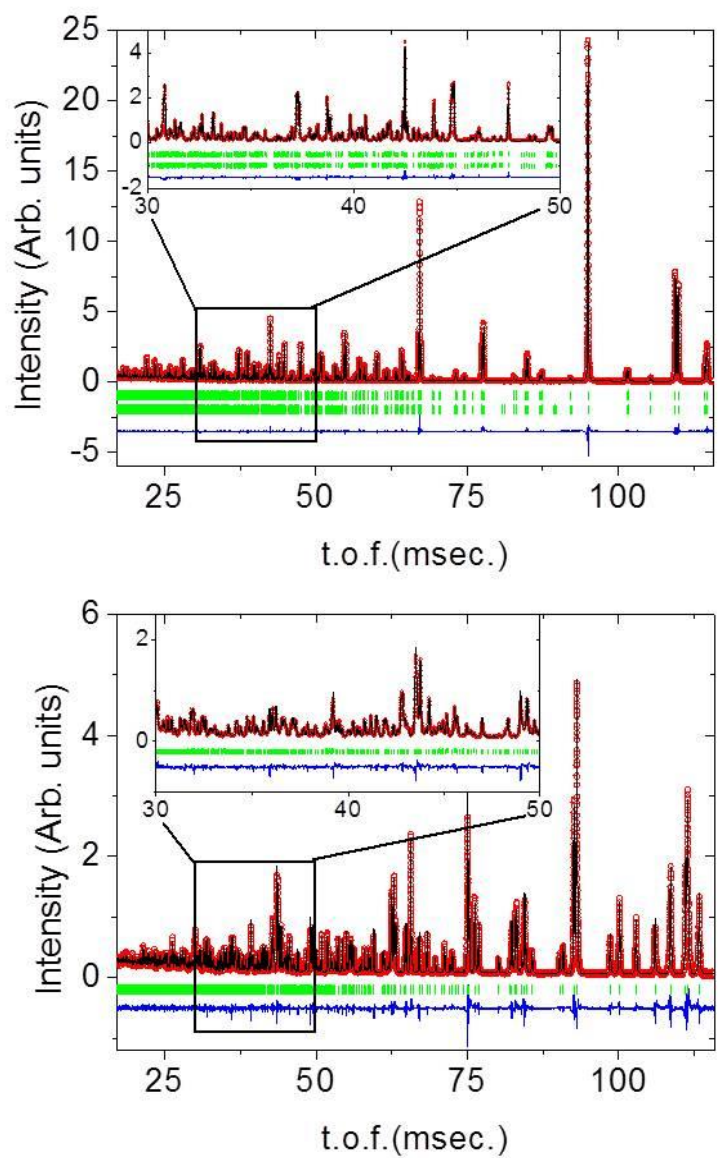

Fig. 3 (Color online) (a) Temperature dependence of lattice constants: a-axis (rectangle), b-axis (triangle), and c-axis (circle), and unit cell volume for $SrRuO_3$ (filled inverse triangle) and $CaRuO_3$ (open inverse triangle). The lines are theoretical curves of the Debye-Grüneisen formula using parameters given in the text. (b) Temperature dependence of three different Ru-O bonds: short in-plane Ru-O1 (circles), long in-plane Ru-O1 (squares), and Ru-O2 (triangles) for $SrRuO_3$ (filled symbols) and $CaRuO_3$ (open symbols) and the vertical dashed line marks $T_c$ with an insert of $RuO_6$ octahedron. Glazer angles estimated based on the experimental data for (c) $SrRuO_3$ and (d) $CaRuO_3$.

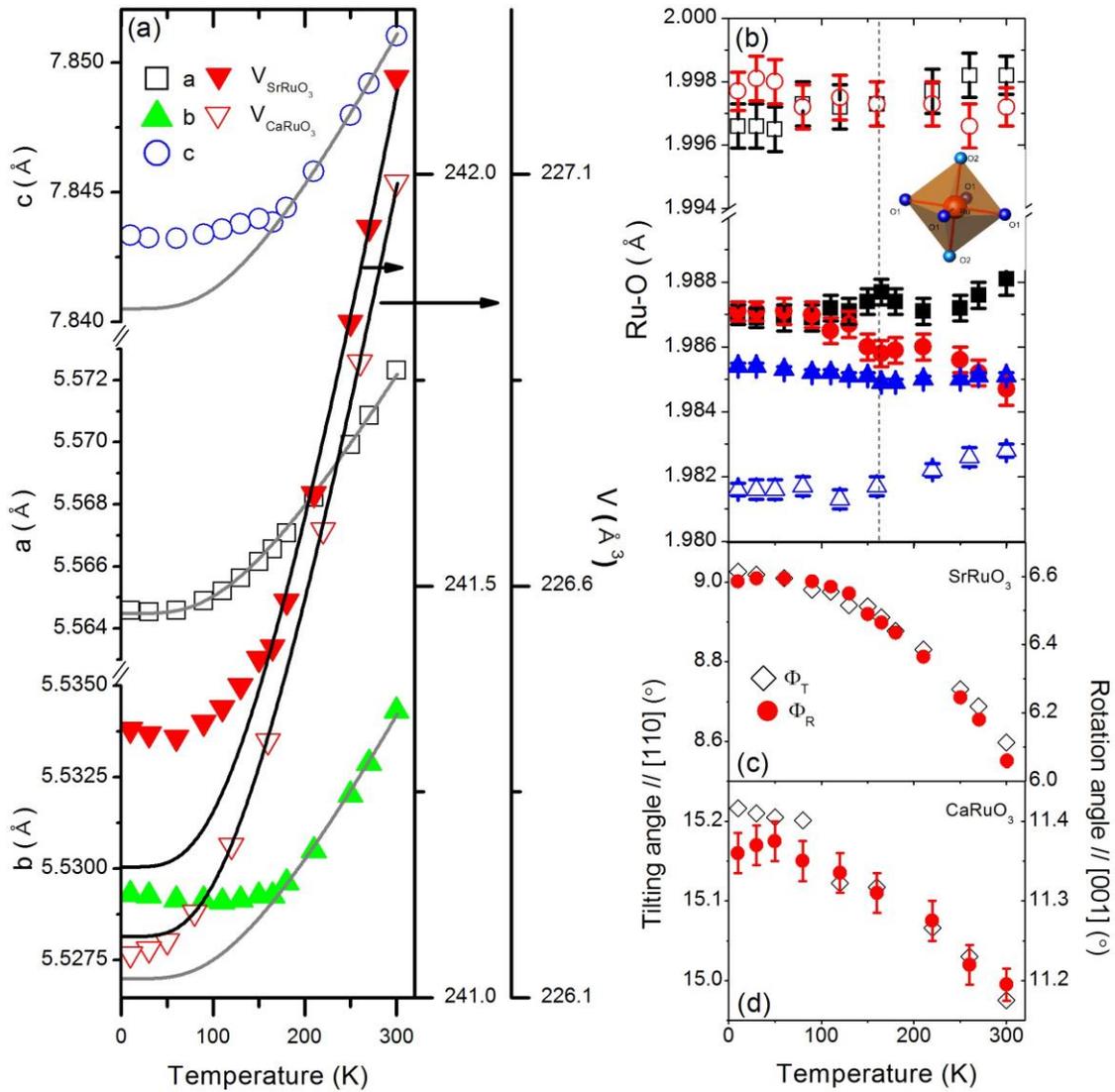

Fig. 4 (Color online) This depicts the definition of the Glazer rotation ($\Phi_R$) and tilting ($\Phi_T$) angles. The rotation angle along the [001] axis is defined as $\Phi_R=(180-\Omega)/2$ while the tilting angle along the [110] axis is given by $\Phi_T =(180-\Theta)/2$.

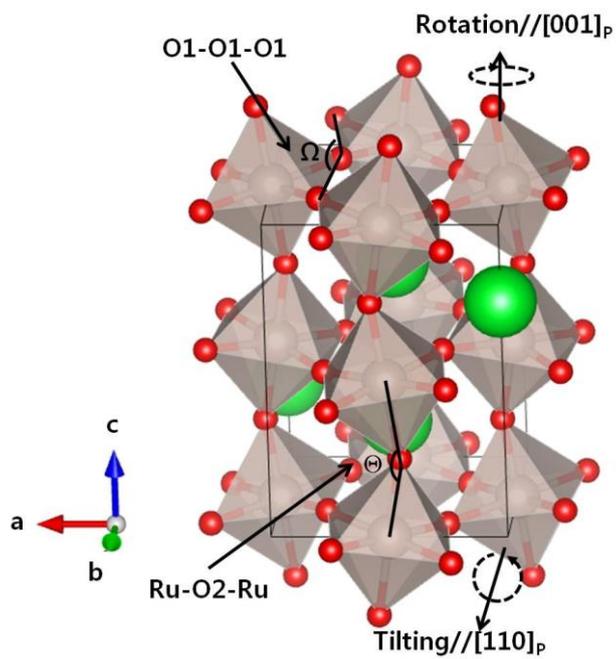

Fig. 5 (Color online) (Left) The ab-plane projection of both $SrRuO_3$ and $CaRuO_3$ with the following values of O1-O1 bond lengths and O1-Ru-O1 bond angles: for $SrRuO_3$ ($CaRuO_3$) $a_{oct}$ is 2.8355 (2.8301) and $b_{oct}$ is 2.7827 (2.8203) Å; (O1-Ru-O1)$_a$ bond angle is 91.08 (90.20)° and (O1-Ru-O1)$_b$ bond angle is 88.92 (89.80)°; (O1-O1-O1)$_b$ bond angle is 167.88 (157.61)°. (Right) Comparison of a and b lattice constants with theoretical b-axis lattice constants calculated in three different methods as discussed in the text. The lines represent our fitting results using the Debye- Grüneisen formula.

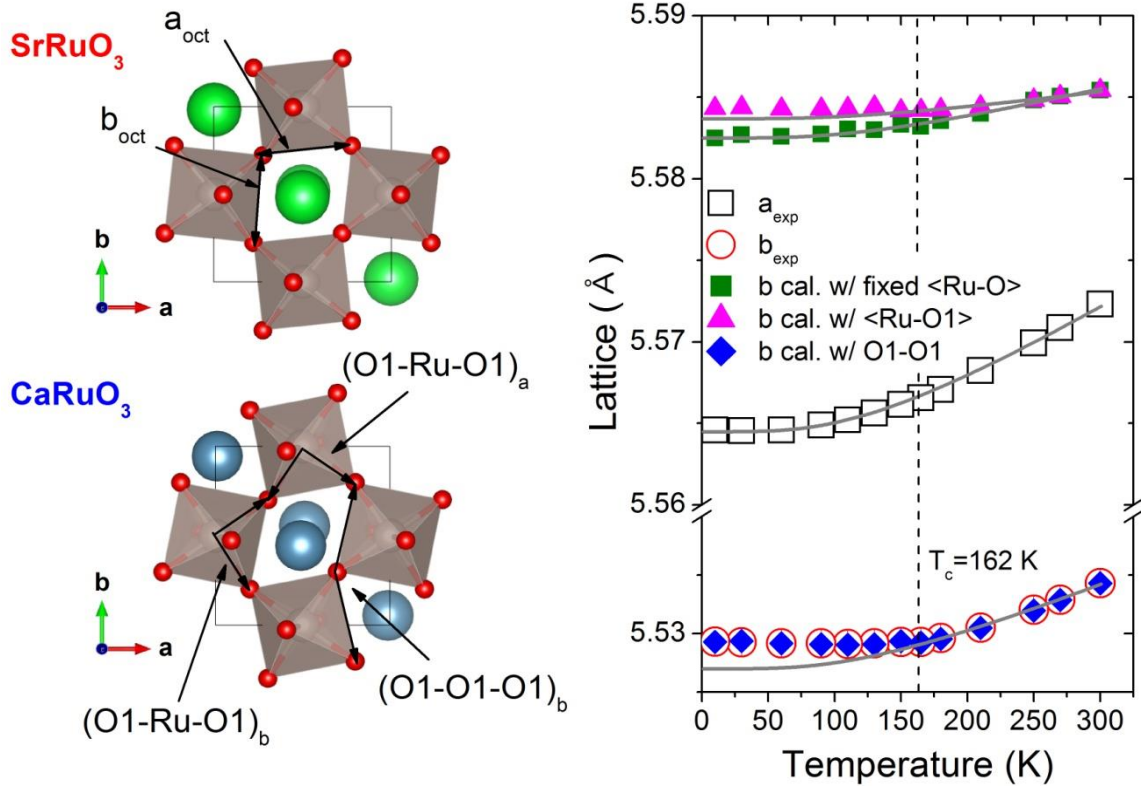

Fig. 6 (Color online) Comparison of the experimental lattice constants (circle) and theoretical values calculated without deformation (rectangle) and with full in-plane deformation (diamond) of $RuO_6$ as discussed in the text. The lines represent our Debye-Grüneisen fitting results of the experimental data. Error bars of the experimental data are smaller than the symbol size.

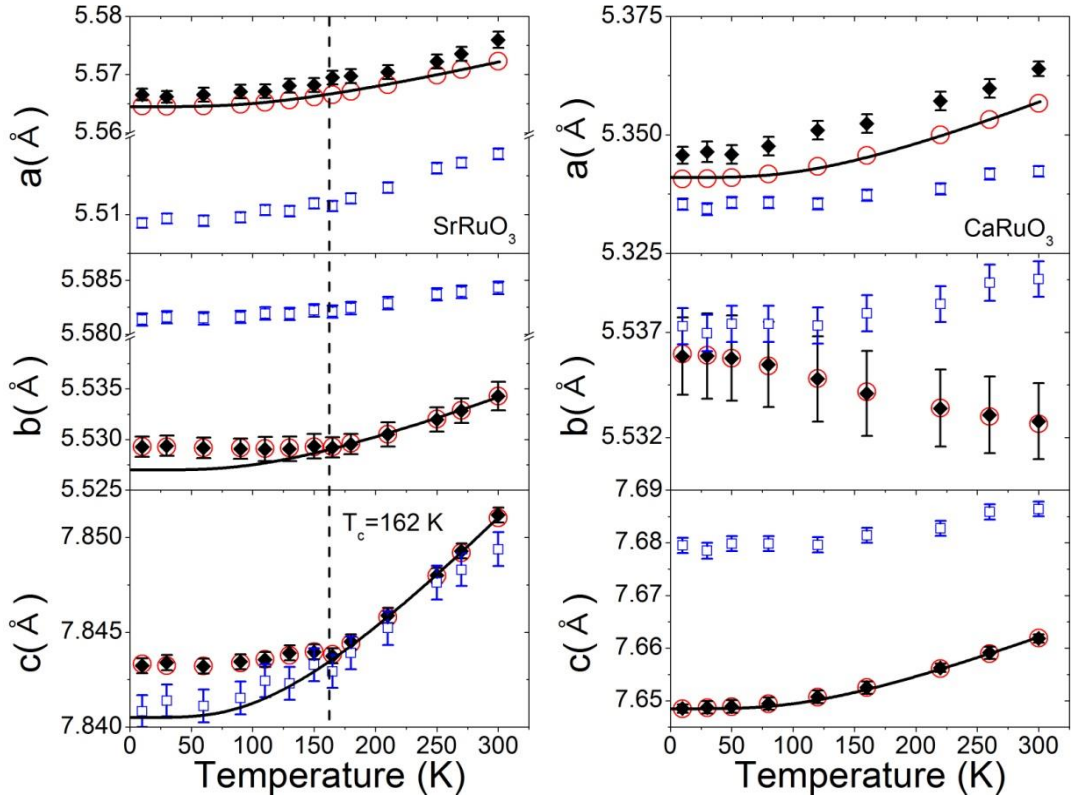

Fig. 7 (Color online) (a) Temperature dependence is shown of the average <Ru-O> bond length for both $SrRuO_3$ (diamond) and $CaRuO_3$ (circle) together with data for $La_{0.75}Ca_{0.25}MnO_3$ (triangle) taken after Ref. 20. The horizontal dashed line indicates an average value in the paramagnetic phase of $SrRuO_3$, while the solid line underneath the data points of $CaRuO_3$ is a theoretical line using the Debye-Grüneisen formula. The excess Ru-O bond length of $SrRuO_3$ above the dashed horizontal line is defined as $\Delta d_{<Ru-O>}$. (b) Scaling plot for two physical quantities of structural and magnetic origin: the ordered moment ($\mu_{ord}$) measured by neutron diffraction and the changes in the <Ru-O> bond length ($\Delta d_{<Ru-O>}$). For the sake of better presentation, we have normalized the physical quantities against at the values at the lowest temperature and they are plotted as a function of reduced temperature, i.e. $\frac{T_C - T}{T_C}$. The line is a theoretical curve of mean field type. (c) Normalized theoretical bandwidth of the three compounds as discussed in the text.

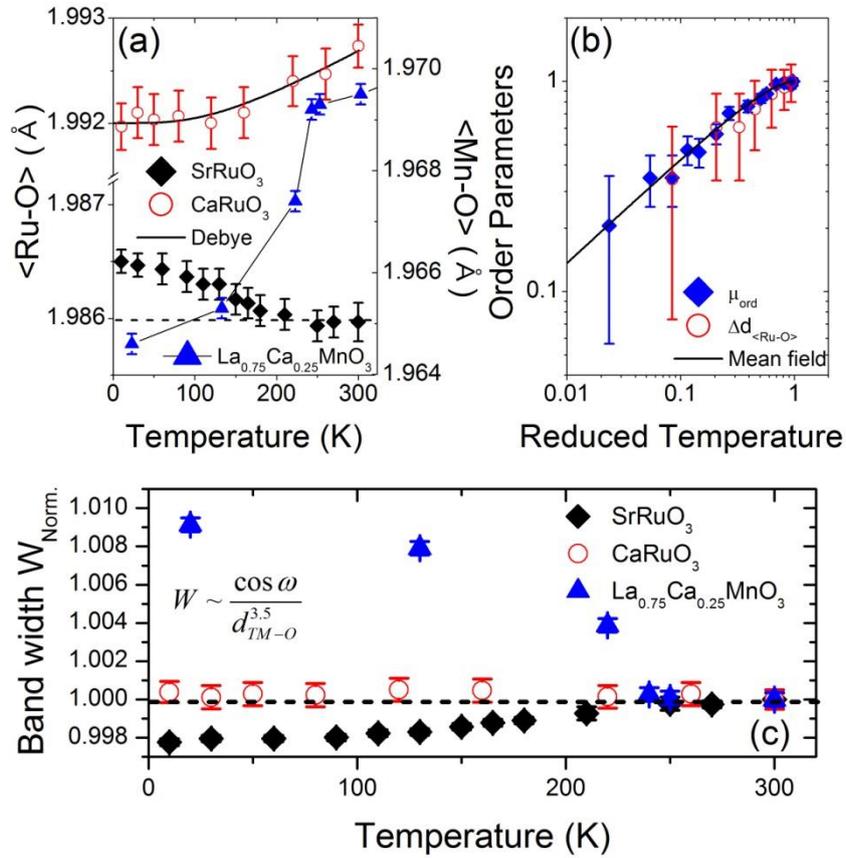

Table 1 Summary of Rietveld refinement results. Atomic parameters are given for SrRuO$_3$ and CaRuO$_3$ as determined from high resolution neutron diffraction patterns at three respective temperatures. The crystal symmetry is orthorhombic *P b n m* with the following atomic positions: Sr and Ca at 4c (x, y, 0.25); Ru at 4b (0.5, 0, 0); O(1) at 8d (x, y, z); O(2) at 4c (x, y, 0.25).

| Structure parameter | SrRuO$_3$ | | | CaRuO$_3$ | | |
|---|---|---|---|---|---|---|
| | 10K | 165K | 300K | 10K | 160K | 300K |
| a (Å) | 5.56458(1) | 5.56656(1) | 5.57231(1) | 5.34064(2) | 5.34567(2) | 5.35661(2) |
| b (Å) | 5.52931(1) | 5.52927(1) | 5.53431(1) | 5.53647(2) | 5.53468(2) | 5.53315(2) |
| c (Å) | 7.84333(2) | 7.84384(2) | 7.85103(2) | 7.64851(3) | 7.65253(3) | 7.66191(3) |
| Volume (Å$^3$) | 241.326(1) | 241.426(1) | 242.117(1) | 226.153(2) | 226.412(2) | 227.090(1) |
| | | | | | | |
| Sr/Ca x | -0.00281(8) | -0.00241(10) | -0.00171(11) | -0.01189(22) | -0.01135(25) | -0.01095(21) |
| Sr/Ca y | 0.02105(6) | 0.01946(8) | 0.01665(11) | 0.05697(17) | 0.05652(19) | 0.05466(16) |
| O1 x | 0.72132(6) | 0.72185(7) | 0.72363(8) | 0.69794(12) | 0.69824(13) | 0.69888(11) |
| O1 y | 0.27903(6) | 0.27877(7) | 0.27717(8) | 0.29830(11) | 0.29829(12) | 0.29805(10) |
| O1 z | 0.02877(4) | 0.02847(5) | 0.02753(6) | 0.04916(8) | 0.04899(8) | 0.04854(7) |
| O2 x | 0.05583(9) | 0.05513(10) | 0.05317(12) | 0.09333(16) | 0.09266(18) | 0.09170(15) |
| O2 y | 0.49587(9) | 0.49653(11) | 0.49706(14) | 0.47318(16) | 0.47336(18) | 0.47366(15) |
| Agreement factors(%) | Rwp=7.51 | 8.06 | 8.58 | 12.0 | 12.3 | 10.1 |
| | Rp=5.33 | 5.78 | 6.22 | 8.40 | 8.92 | 7.42 |